# Unique supramolecular assembly through Langmuir – Blodgett (LB) technique

S. A. Hussain[1]*, B. Dey[1] and N. Mehta[2]

[1]Thin film and Nanoscience Laboratory, Department of Physics, Tripura University, Suryamaninagar 799022, Tripura, India

[2]Physics Department, Institute of Science, Banaras Hindu University, Varanasi 221005, India

**Abstract:** The Langmuir-Blodgett (LB) technique is a way of making supra-molecular assembly in ultrathin films with a controlled layered structure and crystal parameter, which have many envisioned technological applications for optical and molecular electronic devices as well as signal processing and transformation. Probably LB technique is the best method to manipulate materials at molecular level and provides a scope to realize the molecular electronics in reality. In this review article, this important film preparation has been discussed, highlighting its application potential.

**Corresponding author Email:** sa_h153@hotmail.com, sahussain@tripurauniv.in

**Keywords:** Langmuir-Blodgett films, thin films, sensor, opto-electronic device

# 1. Introduction

Thin films with carefully designed structures and properties have attracted a great deal of interest in recent years[1-6] due to their potential applications in a number of different fields, such as sensors[7-8], detectors[1-6], surface coating[1-6], optical signal processing[9-11], digital optical switching devices[12-14], molecular electronic devices[15-20], nonlinear optics and models mimicking biological membranes[21-23]. These applications require, in general, well ordered films consisting of molecules with specific properties, carefully aligned with respect to each other and the substrates possessing high degree of stability to thermal and chemical changes. Langmuir-Blodgett (LB) technique is a powerful tool in creating carefully controlled supramolecular structures of organized molecular assemblies[24-29], which have their potential applications in above mentioned areas. The bulk properties of the materials incorporated into LB films can be controlled by the organization of molecules in the films and also by changing various LB parameters. The possibility to synthesize organic molecules almost without limitations and with desired structure and functionality in conjunction with LB film deposition technique enables the production of electrically, optically and biologically active components on a nanometer scale. Perhaps the LB technique is the most suitable methods for the realization of nano to micro order thin films of organized system with wide range of application potential[5].

In this paper an overview of the LB technique and its application potential have been reviewed.

# 2. Langmuir–Blodgett (LB) methodology

The Langmuir – Blodgett (LB) technique offers the possibility to obtain highly ordered well defined and controlled mono / multilayers and realizes the construction of ultimate molecular architectures which allow the study of physical phenomenon on a molecular level.

Typically LB compatible materials are amphiphilic molecules having two distinct regions, which are a hydrophilic head group (water loving) and a hydrophobic tail group (water hating) as shown in figure 1a. They must be soluble in organic non polar and water immiscible solvents. Amphiphilic molecules form insoluble monolayer at air-water interface. Long chain fatty acid, lipid molecules etc are the example of LB compatible materials.

Initially minute amount (few drops) of dilute solution of amphiphilic materials dissolved in water immiscible solvent such as chloroform are spread onto the air-water interface on the trough of the LB instrument. The molecules rapidly spread all over the available surface area. After solvent evaporation almost one molecule thick layer of the amphiphile remained at the interface, with the head groups immersed in the water and the tail group lying outside (figure 1b). Normally the amount of molecules (concentration) is kept sufficiently low so that initially the molecules remained far from each other and exert very little force on each other. Under these circumstances, the interaction between the molecules is very low (no lateral adhesion) and the resulting floating monolayer can be considered as a two – dimensional gas due to the large inter molecular distances (figure b & c).

Idea about the surface behaviour as well as stability of the floating monolayer has been estimated by recording surface tension. Presence of molecules at air – water interface affects the liquid surface tension. In the gaseous state, the monolayer has relatively little effect on the surface tension. As the available surface area of the monolayer is reduced by a computer controlled barrier minutely, the intermolecular distance decreases and the surface tension decrease further. The amphiphilic molecules start to interact with each other. The force exerted by the film per unit length corresponds to a two dimensional analogue of a pressure, called surface pressure ($\pi$). Typically, this surface pressure ($\pi$) is equal to the decrease in surface tension of air – water interface due to the presence of monomolecular film and can be expressed as:

$\pi = \gamma_0 - \gamma$

where $\gamma_0$ is the surface tension of the empty air – water interface and $\gamma$ the surface tension of the same in the presence of a monolayer.

During the compression of the monolayer, self organization of the molecules occur and the monolayer undergoes several phase changes analogous to the three dimensional gaseous, liquid and solid states. In solid state the molecules finally form a compact and well ordered two dimensional monolayer at air – water interface (figure 1b). These changes in monolayer phases during compression are monitored by measuring the surface pressure ($\pi$) as a function of the area available to each molecule (area per molecule, expressed in nm$^2$) and the corresponding plot is known as surface pressure – area ($\pi$ –A) isotherm. Schematic of a typical $\pi$ –A isotherm curve is shown in figure 1c. In LB technique the $\pi$ –A isotherm can be considered as two dimensional finger print to have idea about the thermodynamic behaviour of the floating monolayer[30,31].

It must be mention this context that the shape of the isotherm greatly depends on subphase temperature, pH, hydrocarbon chain length and the head group. Depending on these factors different transition phases can be observed[32]. As a whole the π –A isotherm provide information on the monolayer stability at the air – water interface, the reorientation of the molecules in the two dimensional system, and the existence of phase transitions and conformational transformations[31]. For further discussions on pressure – area isotherms, the reader can refer to different books and reviews dedicated to Langmuir and Langmuir–Blodgett films[31–33].

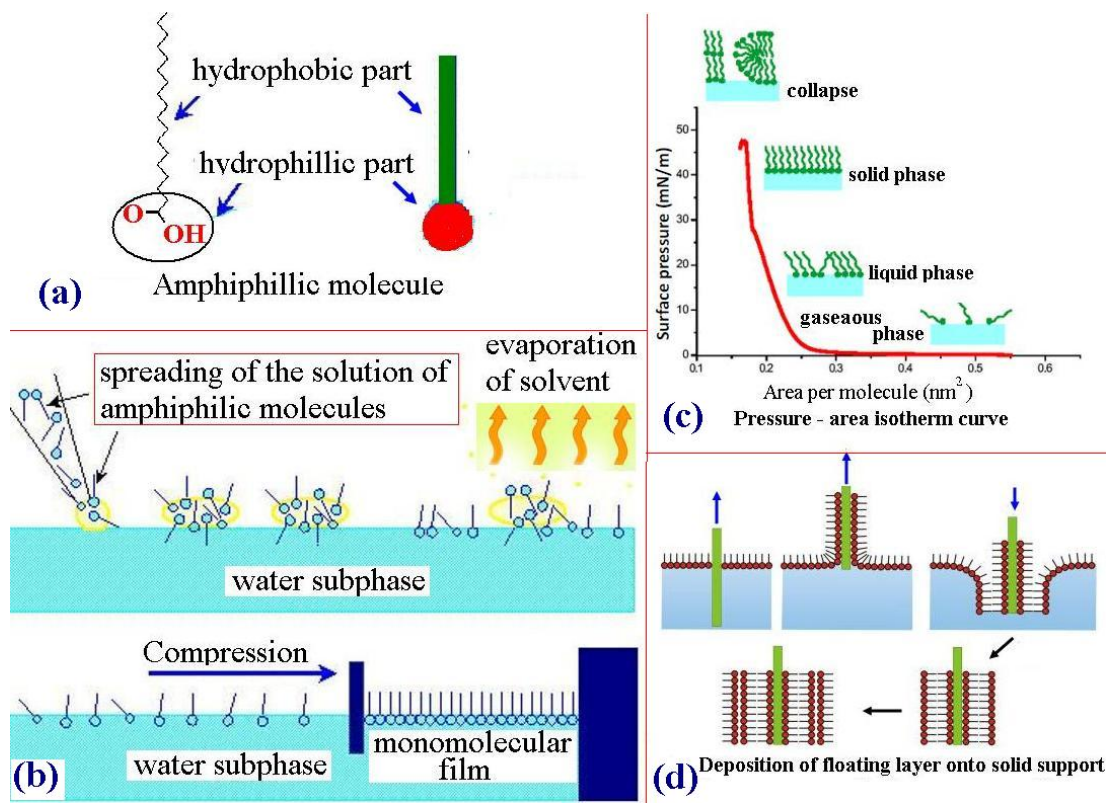

Figure 1: Langmuir – Blodgett (LB) technique

| Table 1: Factors governing the monolayer phase behavior |
|---|
| **Van der Waals interaction:**<br>- Short range forces: ($\Gamma_{VDW} \propto r^{-6}$)<br>- Directly proportional to the number of carbon atoms in the hydrocarbon chain for the saturated hydrocarbon<br>- Depends on the presence of double bonds or substituents like methyl group in the chain |
| **Electrostatic interaction:**<br>- Ion – Ion ( $\Gamma_{ii} \propto r^{-1}$ )<br>- Ion – Dipole ( $\Gamma_{id} \propto r^{-2}$ )<br>- Dipole – Dipole ( $\Gamma_{dd} \propto r^{-3}$ ) |
| **The subphase composition:**<br>- pH of the subphase<br>- Temperature of the subphase<br>- Type / nature of materials |

A mono or multilayer film of the floating monolayer can be deposited onto solid support (often called substrate) at a desired constant surface pressure. There exists a number of different ways in which the interfacial monolayer can be transferred. In Langmuir–Blodgett technique [31,33] this is done by the vertical movement of a solid substrate through the monolayer at air –interface (Figure 1d). Based on the nature (hydrophilic or hydrophobic) of the substrate, the first monolayer will be transferred as the substrate is either raised or lowered through the interfacial film. Subsequently, successive depositions of single layers occurred during each traversal of the substrate through the monolayer film at air – water interface. Such a deposition mode is called Y-type. This leads to a stack (multilayer) in head-to-head and tail-to-tail configuration[31]. Y type deposition is very often used in LB technique. However, in certain cases the floating monolayer is transferred only during the immersion (X-Type deposition) or the emersion (Z-type deposition) of the substrate. LB film deposition typically depends on the pH & temperature of the subphase, humidity, deposition speed, deposition pressure, layer number and contact angle between the substrate and floating layer during deposition[34-40].

| Table 2: The substrates |
|---|
| Monolayer can be transferred to many different substrates according to need viz – <ul><li>Glass , quartz, mica etc</li><li>Aluminum , tin and their oxide</li><li>Silicon wafer, Gallium arsenide wafer</li><li>Gold, Silver</li><li>Cleaning of substrates is very important</li></ul> |

| Table 3: Why LB film studies are important? |
|---|
| <ul><li>Homogeneous deposition of the monolayer over large areas</li><li>Multilayer structures with varying layer composition</li><li>Assembly of individual molecules into 2D & 3D systems</li><li>Control of film structure & monolayer thickness at the molecular level</li><li>Almost any kind of solid substrate can be used</li><li>Unique blend of Physics, chemistry & molecular engineering</li><li>Understanding of structure property relations</li></ul> |

Ideally the LB technique offers the possibility to control and manipulate each step of the LB film preparation precisely. The main advantage of LB deposition is the attainment of a molecular arrangement perfectly organized at the air – water interface, which can be maintained during the transfer onto the solid support by optimizing all the parameters.

## 3. Characterizations of Langmuir–Blodgett (LB) films:

| Experimental technique | Information extracted |
|---|---|
| Brewster Angle Microscopy (BAM) | In-situ study of thin films at various phases at interfaces. It is Sensitive to the surface density and to the anisotropy of domains in monolayer. |
| Fluorescence Imaging Microscopy (FIM) | In-situ domain structure study (micrometer range) |
| Surface potential | Polarization, orientation. |
| Infrared reflection absorption spectroscopy, Attenuated total reflection – Fourier transform infrared (ATR-FTIR) | Hydrocarbon chain packing & conformation, degree of ionization of the head groups, H-bonding, chemical and structural changes, molecular orientation. |
| Ellipsometry | Refractive index & thickness measurement ($\sim 2 \text{A}^0$) |
| X-ray diffraction / reflection<br>Grazing incidence – X-ray diffraction | Inter layer spacing, in-plane lattice structure etc<br>In-plane lattice structure of molecular assembly and hydrocarbon chain tilt. |
| Neutron diffraction | Inter layer spacing. |
| UV-Vis absorption spectroscopy | Electronic transition & orientation. |
| Raman spectroscopy & Surface-enhanced Raman spectroscopy (SERS) | Identification & orientation, conformation of alkyl chains and head groups, molecular interactions with LB films. |
| Optical harmonic generation ($2^{nd}$ & $3^{rd}$ order) | Non-linear coefficient, orientation. |
| Optical microscopy<br>Fluorescence near-field scanning optical microscopy | In-plane structural data.<br>Molecule orientation, lipid domain morphology, grain boundaries, microcollapsed region (lateral resolution of 0.1 μm). |

| Scanning electron microscopy (SEM) Transmission electron microscopy (TEM) | Surface morphology, domain structure, patterns, pinholes and defects (in-homogeneous crystalline domains, micro-collapse etc) (resolution of 50 nm). |
|---|---|
| Scanning tunnelling microscopy (STM) Atomic force microscopy (AFM) | Surface topography, domain morphology, visualisation of defects (grains, pinholes, lateral heterogeneity, disclinations etc), sub-molecular packing (lateral resolution of 0.2 nm). |
| X-ray photoelectron spectroscopy (XPS) or electron spectroscopy for chemical analysis (ESCA) | Quantitative analysis of chemical composition of the film surface. |
| I – V & C – V, resistivity etc measurement | Electrical behaviour. |
| For further details, the reader can refer to different related books and review papers.[31-33,47-67]. | |

## 4. Langmuir – Blodgett (LB) films and Molecular Electronics:

Of late Molecular electronics has been emerged as an important technology for the 21st century[1-6,7-8]. The LB technique provides one of the few methods of preparing organized molecular assemblies which are the pre-requisites for molecular electronic devices. LB technique is perhaps the earliest example of "supramolecular assembly", providing the opportunity to exercise molecular level control over the structure of organic thin films. It is the LB technique which triggers us to dream about the molecular electronics where organic molecules will perform an active function in the processing of information and in transmission and storage. LB films have been utilized for applications that include electronics, optics, microlithography, and chemical sensors, as well as biosensors or biochemical probes[31,33]. The following sections of this article describe some typical application involving LB films.

## 5. Historical review of Langmuir – Blodgett films:

The interesting effects of oil on water were known to the ancients. However, the first attempt to place the subject on a scientific basis was made by Benjamin Franklin, in the eighteenth century[41]. Lord Rayleigh[42] first suggested that the maximum extension of an oil film on water consists of a layer of one molecule thick. Agnes Pockel demonstrated that this Oil film could be controlled by movable barrier[43]. Her simple apparatus later became the model for what is now termed as Langmuir trough. Extensive monolayer theory was developed by I. Langmuir[44]. He was awarded Nobel prize for his experiments suggesting that adsorbed films of great stability could be formed in which single layers of atoms were bound by the underlying surface. However, the first detailed description of sequential monolayer transfer was given several years later by K. Blodgett[45]. These built up monolayer assemblies are now called Langmuir-Blodgett (LB) films. After the pioneering work done by Langmuir and Blodgett, it took almost half a century before scientists all around the world started to realize the opportunities of this unique technique. It was Khun and his collaborators who revived and demonstrated the potential of the field[46]. His work stimulated interest in the topic in Europe, USA and Japan and now throughout the world. The first international conference on Langmuir-Blodgett (LB) films was held in 1979 and since then the use of this technique is increasing among the scientists throughout the world working on various different field of research. Extensive studies on LB films are now being going on. Various review articles has been published time to time since the revival of the interest in research related to different aspects of LB films. Readers may go through the list of review article given in the following table for further information.

| Topic of the review | Ref. No. |
|---|---|
| An applied science perspective of Langmuir-Blodgett films | 47 |
| The Physics of Langmuir – Blodgett films | 48 |
| Langmuir – Blodgett films | 2 |
| Langmuir-Blodgett Films - Properties and Possible Applications | 49 |
| Langmuir–Blodgett films: molecular engineering of non-centrosymmetric structures for second-order nonlinear optical applications | 50 |
| Supramolecular organization of highly conducting organic thin films by the Langmuir–Blodgett technique | 51 |
| Mapping molecular orientation and conformation at interfaces by surface nonlinear optics | 52 |
| Surface dilational moduli of polymer and blended polymer monolayers spread at air-water interfaces | 65 |
| Interactions of bioactive molecules & nanomaterials with Langmuir monolayers as cell membrane models | 66 |
| Phase transitions in polymer monolayers: Application of the Clapeyron equation to PEO in PPO–PEO Langmuir films | 67 |

5. **Research and Applications:**

Modern studies of LB films fall largely into two areas – (i) detailed fundamental studies of the physical nature and structure of Langmuir monolayers and LB films and (ii) applications towards passive and active components of devices.

**5.1 Fundamental research:**

Since the discovery the technique, LB technique and LB films are used as model systems for a broad range of scientific research[2,47-54]. Most of the current research in this area derives inspiration from the pioneering work of Hans Kuhn[68,69] who, in the 1960s, moved away from studies on fatty acids and other simple amphiphiles and used LB technique to control the position and orientation of functional molecules within complex assemblies.

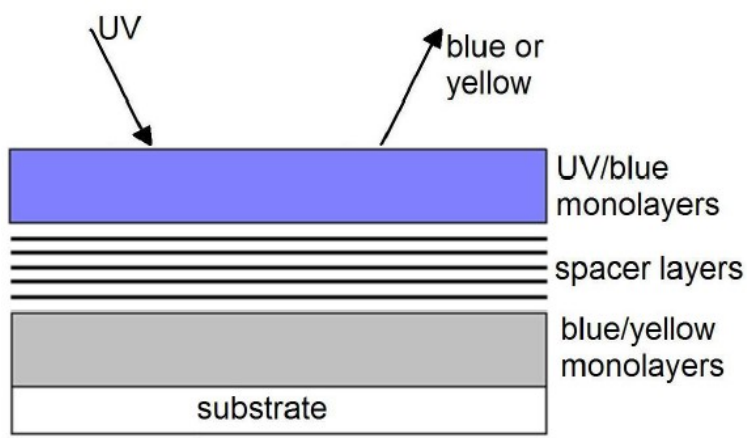

Figure 2: Kuhn's experiment of two blocks of different fluorescent layers, separated by a block of normal "spacer" layers.

One example from Kuhn's work is shown in figure 2. A block of fluorescent LB films that absorbs blue light and emitting in the yellow is separated from another block of different fluorescent LB films that absorbs in the UV range and emitting in the blue, by a block of fatty acids (spacer layer). He demonstrated that assuming a UV-light is incident on the structure, it's possible to have either blue emission or yellow emission from the structure. The overall emission of the structure has thus changed from blue to yellow when the block of spacer layers is thin enough. Here energy transfer between the layers occurred due to the quantum mechanical tunneling, where the spacer layers can control the energy transfer process.

LB technique can be used to mimic natural biomembrane artificially[56,62-63,70-71]. Accordingly LB system has been used in biological research in a variety of ways[55,64,66,72-73]. LB films of certain materials have been used to immobilize an enzyme or a protein, which can trap and bind with some ions or molecules that change some measureable property of the film[56,62,64]. Accordingly the properties of those enzymes and proteins can be studied[74-76]. Lipid-containing artificial tears are used to restore the lipid layer of the tear film. Recently LB technique has been used to investigate the lipid layers containing artificial tears[77].

The biomineral calcium oxalate monohydrate (COM) is a major component of urinary stones. An understanding of the processes that lead to COM precipitation in the urinary tract can lead to the prevention or treatment of urinary stone disease. LB technique has been used to prepare LB films of phospholipids to model domains in biological membranes for studies of COM precipitation and adhesion. From the studies in formation about how the chemical and organizational properties of the model membrane affect the formation of COM crystals at the membrane interface can be explored[78-79]. Lots of efforts have been given to investigate different kinds of molecular aggregate formed onto LB films[34-38]. It has been shown that this technique is very suitable for preparation of different supramolecular aggregates, where the properties of such aggregates can be tuned by optimizing various parameters during film formation.

**5.2 Applications towards passive and active components of devices.**

The most common application of LB films as the passive layers in several areas of commercial interest includes lithographic photoresist[80-83], lubrication[84-86] as well as enhancing device performances such as surface acoustic waves[5,85,87-89], liquid crystal alignment[74,90-92] etc. For creating devices with smaller dimensions the electron-beam lithography is used in research. It requires a very thin resist for enhancement of resolution due to the pronounced scattering in a resist of normal thickness. LB films have been used as photoresist viz, LB films of fatty acid salts as a positive resist and of the 22-tricosenoic acid as a negative resist have been used with good results[80-83]. Langmuir-Blodgett films have been used as lubricating layers for enhancing the useful life of high density hard disks[84]. LB films have also been used for the lubrication of magnetic tapes[86]. The friction coefficient and wear of the tape has been shown to dramatically reduced by coating the tape with seven layers of barium stearate.

The most significant passive application of LB films could however be for the alignment of liquid crystals. The dipole energy of the underlying substrate mainly governs the alignment of liquid crystals. By tailoring the structure of LB it is possible to control the dipole energy and hence the orientation of liquid crystal. It has been shown that some liquid crystal molecule align spontaneously when sandwiched between two glass plates covered with a single layer of a polymer LB film[90,74]. Acousto-electric devices offer many attractive features for applications as physical and chemical sensors. Surface acoustic wave (SAW) oscillators are of particular importance owing to their high sensitivity. Use of LB films as gas absorbent layers on the surface of SAW devices has been demonstrated[88,89]

It is widely recognized that pyroelectric materials have considerable advantages over narrow band-gap semiconductors, as detectors of infrared radiation. There are several advantages that make LB films particularly attractive candidates for pyroelectric devices. The most important of these is that the sequential deposition of single monolayer enables the symmetry of the film to be precisely defined; in particular, layers of different materials can be built-up to produce a highly polar structure. Secondly, the polarization of an LB film is 'frozen-in' during deposition and it is therefore not necessary to subject the film to a poling process. The third advantage is that the LB technique uses amphiphilic organic materials which possess low permitivities ($\varepsilon$) and the figure of merit for voltage responsivity $p/\varepsilon$ (where p is the pyroelectric coefficient) is expressed to be large. Finally the LB method enables the preparation of much thinner films than usually attained by normal conventional techniques. The pyroelectric coefficients for multilayer acid / amine LB films can be about 10 $\mu Cm^2 K^{-1}$ and depend on the thermal expansion coefficient of the substrate, indicating that there is a significant secondary contribution to the measured pyroelectric response[1,33,93-95]. LB films of poly(vinylidene fluoride trifluoroethylene) showed ferroelectric switching with distribution of switching times several decades wide[96].

Rectifying junctions are the basic elements of many electronic components. Rectifying junctions such as polymer p-n junction and schottky junction based on Langmuir-Blodgett (LB) films have been studied widely[97]. Aizawa et. al.[98] have demonstrated that anion doped polypyrrole / polythiphene can behave as n-type semiconductor. Metal-insulator-semiconductor (MIS) structures were fabricated by vacuum deposition of various metals like indium, aluminium and tin on LB films of cadmium stearate ($CdSt_2$) obtained on polypyrrole films deposited on indium-tin-oxide glass[99]. The value of the dielectric constant of the insulating $CdSt_2$ LB film has been found to be 1.84 which is in good agreement with the value reported earlier. Organic heterostructures have been fabricated by alternating deposition of mono- and multilayers of undoped poly (3-hexylthiophene) and doped polypyrrole prepared by LB technique[100]. Electrical rectifying devices (Zener diode) have been prepared using a two layer configuration, consisting of a $p^+$-doped semiconducting polymer Polypyrrole or poly (3-methylthiophene) layer and a n-type multilayer structure of CdSe and 1,6-hexenedithiol[101].

There is much current interest in the possibility of observing molecular rectification using monolayer and multilayer. This follows the prediction of Aviram and Ratner that an asymmetric molecule containing a donor and an acceptor group separated by a short $\sigma$-bonded bridge (allowing tunneling) should exhibit diode characteristics[102]. There have been many attempts to obtain this effect in LB films[33]. However, the demonstration that the rectification is affected by a change in the film structure (by bleaching)[103] and a recent report of rectifying behaviour with symmetrical gold electrodes[104] suggest that true molecular rectification is achievable.

Experimental and theoretical confirmation of anion-induced dipole reversal in cationic dyes due to molecular rectification have been demonstrated[105]. There is a good review article by R. M. Metzger on the application of Langmuir-Blodgett (LB) films as unimolecular rectifier[106-107].

Another interesting application of the LB technique is the fabrication of thin films of copper phthalocyanine derivatves as field effect transistor (FET). It is well known that pthalocyanine derivatives are very promising organic semiconductor materials due to their chemical and thermal stability. Among phthalocyanine, copper phthalocyanine derivatives have been utilized as organic field effect transistors (OFETs). The FET performances of the LB films of phthalocyanine has been tested by I-V curves acquired from devices operating in accumulation mode. It has been observed that the carrier mobility in LB films, can be improved by the arrangement through a more highly ordered film with improved interaction distance and $\pi-\pi$ interaction throughout the FET channel. Recently an interesting work about a physico-chemical investigation of carboxylic ionophores and phospholipids for application as ion selective field effect transistor (ISFET) has been reported[108].

Cui et. al.[109] demonstrated depletion-mode n- channel organic field effect transistors (OFETs) based on naphthalene-tetracarboxylic-dianhydride (NTCDA) transistor, n-type NTCDA acts as active channel material due to its high mobility of 0.06 cm$^2$V$^{-1}$S$^{-1}$, and p-type conducting polymer Polypyrrole performs as the source and drain. Koezuka et. al.[110] fabricated a FET by using two different kinds of conducting polymers, Polypyrrole and polythiophene. Fabrication of Polymer Langmuir-Blodgett Films Containing Regioregular poly(3-hexylthiophene) for application in Field-Effect Transistor (FET) have also been reported[111]. Thin-Film Transistors based on Langmuir-Blodgett Films of Heteroleptic Bis(phthalocyanine) Rare Earth Complexes has been demonstrated[112]. Organic field-effect transistors based on LB films of an extended porphyrin analogue – Cyclo[6]pyrrole and neutral long-chain TCNQ derivatives have also been reported[113,114].

Organic thin films have been used as the semiconducting layers in FET devices[115]. A significant increase in the carrier mobility has been reported over the last fifteen years using LB films of organic materials[116-118].

| Materials | Mobility (cm$^2$V$^{-1}$s$^{-1}$) |
|---|---|
| Polythiphene | 10$^{-5}$ |
| Polyacetylene | 10$^{-4}$ |
| Phthalocyanine | 10$^{-4}$ – 10$^{-2}$ |
| Thiophene oligomers | 10$^{-4}$ – 10$^{-1}$ |
| Pentacene | 10$^{-3}$ – 3 |
| C$_{60}$ | 0.3 |
| Organometallic dmit complex | 10$^{-1}$ |

Table 1: Field effect carrier mobilities of FET based on organic semiconductors[116-118]

Thin-Film Transistors Based on Langmuir-Blodgett Films of Heteroleptic Bis(phthalocyanine) Rare Earth Complexes have also been studied[119]. Solution processed Langmuir-Schäfer and cast thin films of poly (2,5-dioctyloxy-1,4- phenylene-alt-2,5-thienylene) were also investigated as transistor active layers[120]. The study of their field-effect properties evidences that no transistor behavior can be seen with a cast film channel material. This was not surprising considering the twisted conformation of the polymer backbone predicted by various theoretical studies. Strikingly, the Langmuir-Schäfer (LS) thin films exhibit a field-effect mobility of $5 \times 10^{-4}$ cm$^2$/V.s, the highest attained so far with an alkoxy-substituted conjugated polymer. Extensive optical, morphological, and structural thin film characterizations support the attribution of the effect to the longer conjugation length achieved in the Langmuir-Schäfer (LS) deposited film, likely due to an improved backbone planarity. This study shows that a technologically appealing deposition procedure, such as the LS one, can be exploited to significantly improve the inherently poor field-effect properties of twisted conjugated backbones. This achievement could promote the exploitation for electronic, and possibly sensing, applications of the wealth of opportunities offered by the alkoxy substitution on the phenylene units for convenient tailoring of the phenylene-thienylene backbone with molecules of chemical and biological interest.

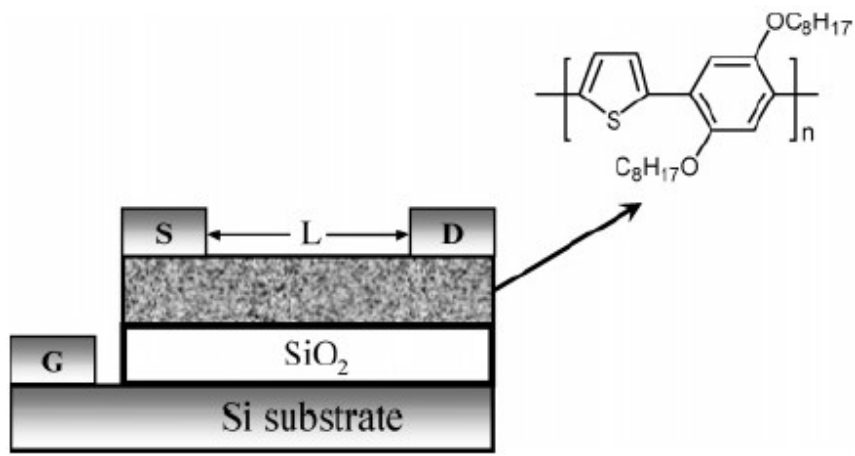

Figure 3: Schematic diagram of a bottom-gate TFT device composed of highly conducting silicon substrate (n-doped, 0.02-1 Ω/cm) coated by 300 nm thick SiO$_2$ thermal oxide ($C_i$ = 10 nF

cm$^2$). Gold source (S) and drain (D) contacts were defined, by thermal evaporation through a shadow mask, directly on the POPT films. The channel length, *L*, is 200 $\mu$m, and the width, *W*, is 4 mm. The gate (G) gold pad was deposited directly on the conducting silicon. The POPT chemical structure is reported on the right[120].

Polymeric light emitting diodes (PLEDs) have shown much interest worldwide since the discovery of electroluminescence (EL) in a thin poly(phenylene vinylene) layer by Friend and coworkers in 1990[121,122]. Conducting polymers have been thought of as potential luminescent materials for replacing inorganic light emitting materials when used in large area, light weight, flexible displays. The main advantage of these materials over conventional luminescent materials are the tuning of wavelength emitted by chemical modification, low operating voltage, flexibility, easy processing, low cost, possibility of making large device and output colours in whole visible spectrum. Several p-doped conducting polymers have been tested in LB films[123-127] and have been used as hole injecting electrodes, like polypyrrole, polythiophene derivatives and polyaniline, which have high work functions, providing low barriers for hole injections.

The relatively ordered nature of arrays of molecules in LB films may be exploited in light emitting structures. For example, the preferential alignment of the molecules can result in polarized light emission[128,129]. LB film deposition can also be used to control the positions of the luminescent species within metal mirror microcavities[130]. To improve the lifetime of OLEDs several experiment on MEH-PPV LB films[131] have been done. More recent work has been focused on hybrid of pyridine and oxadiazole which lead to a further increase in the electron-injection[131].

Many conducting polymers such as polyacetylene, polythiophene, polyindole, polypyrrole, polyaniline etc have been reported as electrode materials for rechargeable batteries[132]. Photoconductivity involves enhancement of the electrical conductivity of the material by the absorption of a suitable photon. It finds wide range of applications in electronics, for example auto brightness control (ABC) circuits in TV sets, camera shutter, car dimmers, street light control, auto gain control in transceivers electrophotography etc. Electrochemically produced polypyrrole LB films (band gap 3.2 eV) after sensitization can be anticipated to exhibit good photoconductivity[133]. Liu et. al.[134] fabricated a polymer based capacitor, using polypyrrole and poly (3,4-ethylenedioxythiphene) poly(styrenesulfonate) as a semiconductor and gate layer. Dielectric polymer, polyvenylphenol, was applied as the insulator to the device. Composite electrodes for supercapacitors were prepared via chemical polymerization[135] of pyrrole on the surface of a porous graphite fiber matrix[135].

**5.3 Sensing application**

Sensor is a device, which provides direct information about the chemical composition of its environment[136]. It consists of a physical transducer and a selective sensing layer. It is invariably provided by a material in which some selective interaction of the species of interest takes place that results in the change of some physical parameters such as electrical current or potential or conductivity, intensity of light, mass, temperature etc. The combination of synthetic chemistry with the molecular engineering capability of the LB technique makes organic multilayered system interesting candidates for sensors. There are numerous physical properties on which LB sensing system can be based. Examples include resistivity changes, electro chemical phenomena, optical effects etc. Electrochemical pesticide sensor based on LB film of cobalt phthalocyanine-anthraquinone hybrid system has been demonstrated[137]. A sensor array made up of nanostructured LB films is used as an electronic tongue capable of identifying sucrose, quinine, NaCl, and HCl at the parts-per-billion (ppb) level, being in some cases 3 orders of magnitude below the human threshold[138]. LB monolayers of tri-n-octylphosphine oxide-capped cadmium selenide quantum dots (QCdSe) onto indium−tin oxide (ITO) coated glass substrate was used to design an electrochemical DNA biosensor for detection of chronic myelogenous leukemia (CML) by covalently immobilizing the thiolterminated oligonucleotide probe sequence via a displacement reaction[139]. Nanostructured PdO LB film shows $H_2$ gas sensitivity at room temperature in a wide range of 30−4000 ppm concentration with fast recovery either on exposure to ambient light or by carrier gas flow. Further studies on the PdO films showed orders of increase in photocurrent on exposure to ambient visible light, suggesting its plausible applications in solar energy conversion[140].

Many different types of organic materials have been used for LB film based gas sensing. These include porphyrins, phthalocyanines and insulating and conducting polymers[141,142]. Electrochemical sensors (i.e. chemiresistors) have been widely investigated (also studies of mass and optical changes) to gain a better understanding of the polymer / gas interactions. Non-conductive polymers (i.e. those not containing $\pi$-bonds) have been used as mass sensors, thermal sensors, optical sensors and dielectric sensors[143,144]. Polypyrrole (PPy) was widely used as various kinds of sensors depending on their transducing mechanism, including mass sensor[145], potentiometric sensor[146-148], potentiometric humidity sensor[145], amperometric sensor[149], amperometric biosensor[150-152], piezoresonance sensors[158] and conductometric sensor[153-157].

Bio sensors are devises capable of retrieving analytical information from the operational environment by utilizing biological components as part of the sensor. Biosensors use biological molecules, mainly enzyme, lipids etc as the recognition elements. Professor Hou,[159] demonstrated a novel biosensor using LB technique to sense the mixtures of odorants in various environmental conditions.

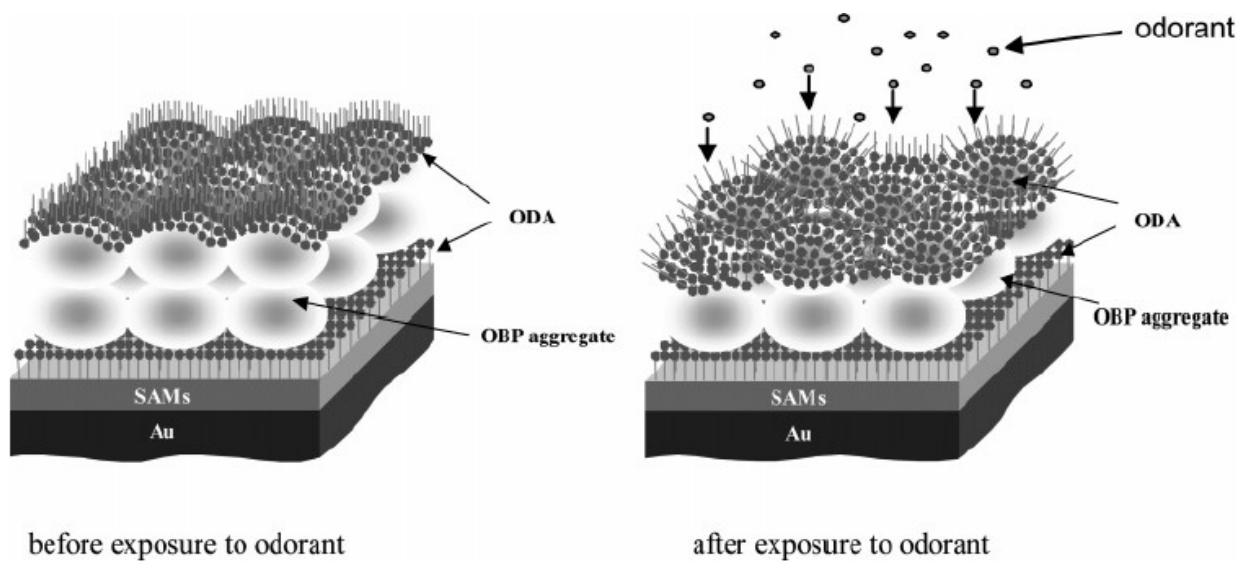

Figure 4. Left: Theoretical model for the two layers of mixed OBP-1F/ODA LB films transferred onto functionalized gold substrate at 35 mN m-1 before exposure to odorant molecules. Right: The model for the two layers of mixed OBP-1F/ODA LB films after exposure to odorant molecules[159].

A novel optical nanosensor using a support bilayer lipid membrane (SBLM) has been recently proposed[160]. In this work LB and Layer-By-Layer (LBL) techniques have been combined to obtain highly ordered nanostructure. This work is particularly significant due to the importance of sensors for biological agents in vivo and/or in vitro. A heptamer linear RGD (acridine-glysine-asparate) containing peptide was covalently attached to a BODIPY (2-(4,4-difluoro-5, 7-diphenyl-4-bora-3a, 4a-diaza-s-dodecanoyl)-1-hexadecanoyl-glycero-3-phospho

ethanolamine, donor) lipid dye and utilized as an optical biosensor. A second BODIPY (4,4-difluoro-5-(2-thienyl)-4-bora-3a,4a-diaza-s-indacene-3-dodec-anoic acid, acceptor) lipid dye was integrated into the SBLM, enabling the signal amplification via a Forster resonance energy transfer (FRET) mechanism. The result indicates the possibility to detect HUVEC at a concentration of 1000 cells ml$^{-1}$. The sensitivity obtained by this method is similar to polymerized chain reaction (PCR) technique methods but less sensitive than flow cytometric techniques[108,161].

A glucose sensor consisting of a conductive polypyrrole membrane and a lipid LB film has been investigated[162]. Professor M. Rikukawa,[163-164], fabricated a similar biosensor comprised of a lipid-modified glucose oxidase and conducting PPy LB film for detecting glucose. T. N. Misra, et. al.[165], prepared a stable monolayer of alcohol dehydrogenase (ADH) enzyme by spreading an aqueous solution of ADH on a water subphase containing stearic acid monolayer. This ADH-stearic acid monolayer has been successfully transferred onto a conducting polypyrrole –coated glass electrode by the LB technique. This LB-immobilized polypyrrole-mediated enzyme electrode can be used as an ethanol sensor. Novel enzyme based micrelectrochemical devices, based on changes in the conductivity of polypyrrole layers, were developed for biosensing of NADH and penicillin[166].

Therefore it is clear that LB technique already showed enormous potential towards both fundamental research and device application. However, main challenges are the development of new sensors and in the production of cheap, reproducible and reliable devices with adequate sensitivities and selectivity.

## 6. Conclusion

Langmuir-Blodgett deposition technique is very simple but very powerful tool towards formation of nano dimensional ultrathin films with desired properties / architectures. Here each film is built from multiple monolayers. Accordingly the thickness and molecular arrangement is controllable at the molecular level. Using LB technique, it is possible to create complex yet precise artificial molecular arrays with predesigned physical and chemical properties. Already the potential of this versatile technique has been tested towards realization of optoelectronic devices. However, LB films have yet to be utilized in today's commercial products. The technique is still expected to be a vital part of future fabrication methods in molecular electronics. Therefore, when molecular electronic and bio-electronic devices will become available, LB films will definitely play a vital role in their realization. Hence, development of LB films for practical / commercial applications is a challenge, requiring an interdisciplinary outlook which neither balks at the physics involved in understanding assemblies of partially disordered and highly anisotropic molecules, nor at the cookery involved in making them. Although LB films cannot be adapted to all purposes, there are signs that with sufficient understanding, their behavior can be optimized for specific technological applications. Therefore it is highly appropriate to make a great stride in these important and promising areas of research, which can provide a conceptual understanding with wide opportunity of technological applications.

**Acknowledgment**

The authors are grate full to DST, Govt. of India, for financial support to carry out this research work through DST project Ref: EMR/2014/000234 and FIST DST project Ref. SR/FST/PSI-191/2014. The authors are also grateful to UGC, Govt. of India for financial support to carry out this research work through financial assistance under UGC – SAP program 2016.